\documentclass[aip,reprint,superscriptaddress]{revtex4-1}

\usepackage{amsmath, amsfonts}
\usepackage{graphicx}
\usepackage[utf8]{inputenc}

\begin{document}
\title{A low-temperature external cavity diode laser for broad wavelength tuning}
\author{William G. Tobias}
\author{Jason S. Rosenberg}
\author{Nicholas R. Hutzler}
\author{Kang-Kuen Ni}
\affiliation{\mbox{Department of Chemistry and Chemical Biology, Harvard University, Cambridge, Massachusetts, 02138, USA}}
\affiliation{\mbox{Department of Physics, Harvard University, Cambridge, Massachusetts, 02138, USA}}
\affiliation{\mbox{Harvard-MIT Center for Ultracold Atoms, Cambridge, Massachusetts, 02138, USA}}
\date{\today}

\begin{abstract}
We report on the design and characterization of a low-temperature external cavity diode laser (ECDL) system for broad wavelength tuning. The performance achieved with multiple diode models addresses the scarcity of commercial red laser diodes below 633 nm, which is a wavelength range relevant to spectroscopy of many molecules and ions. Using a combination of multiple-stage thermoelectric cooling and water cooling, the operating temperature of a laser diode is lowered to $-64^\circ$C, more than $85^\circ$C below the ambient temperature. The laser system integrates temperature and diffraction grating feedback tunability for coarse and fine wavelength adjustments, respectively. For two different diode models, single-mode operation was achieved with 38 mW output power at 616.8 nm and 69 mW at 622.6 nm, more than 15 nm below their ambient temperature free-running wavelengths.  The ECDL design can be used for diodes of any available wavelength, allowing individual diodes to be tuned continuously over tens of nanometers and extending the wavelength coverage of commercial laser diodes.
\end{abstract}

\maketitle

\section{Introduction}
Diode lasers are compact, robust optical sources with a broad range of applications in physics research. However, a major limitation is the incomplete wavelength coverage of commercial laser diodes. The output wavelength is determined by properties of the semiconductor material inside the laser cavity, and for any given atomic or molecular transition there may not be a corresponding diode available. For example, while many commercial diode products based on AlGaInP semiconductors exist for red wavelengths above 630 nm, few exist for the wavelengths below. Applications for lasers of these wavelengths (often in combination with frequency doubling) include slowing, cooling, trapping, and quantum-state-preparation of several different species of diatomic molecules \cite{Loh2012,Loh2011, Hummon2011, Zhelyazkova2014} as well as laser cooling of beryllium ions for quantum information processing \cite{Monroe1995,Ball2013}. Techniques that have been previously employed to construct lasers in the 620 nm range include frequency doubling\cite{Harkonen2007,Konttinen2014}, custom-fabrication of semiconductor materials\cite{Shimada2011}, and cryogenic cooling\cite{Bohdan2008}, but these methods are typically costly and highly dependent on the specific final lasing wavelength. Alternatives to diode lasers include dye lasers, which operate at high power for many red wavelengths\cite{Hammond1979} but are maintenance-intensive, and optical parametric oscillators, which are broadly tunable across the visible spectrum\cite{Kawasaki1994} but are expensive to manufacture.

Various other strategies are used to extend the usable range of commercial laser diodes. A laser diode can be ``pulled'' to a shorter or longer wavelength by introducing feedback light of the desired wavelength to the laser cavity. External cavity diode lasers (ECDLs) use diffraction gratings to provide feedback to the laser diode and can typically pull the wavelength by a few nanometers around the free-running wavelength. The output wavelength can also be tuned by adjusting the temperature of  laser diodes. Laser diode wavelengths typically tune between 0.15-0.25 nm/$^\circ$C due to temperature-dependent changes in the gain profile of the semiconductor diode materials and the length of the laser cavity \cite{Wieman1991}. 

Compact cooling systems for laser diodes and other experimental systems often utilize thermoelectric coolers (TECs) as the primary cooling mechanism\cite{Fletcher2004,Carter1991}. TECs are semiconductor devices that transfer heat from one ceramic or metal-coated ceramic face (the ``cold side'') to another (the ``hot side'') and can maintain temperature gradients greater than $120^\circ$C in optimal, low heat load conditions. High-performance TECs typically consist of a stack of multiple-layer TECs. Previous designs for temperature-tunable lasers reached minimum temperatures of approximately $-40^\circ$C using multiple-stage TECs and insulation. In combination with an external cavity, these systems have achieved large wavelength changes with narrow linewidths at moderate output power \cite{Fletcher2004,Ball2013}.

In this work, we present a temperature-tunable ECDL that achieves an operating temperature of $-64^\circ$C, which extends the tuning range of a red laser diode by more than 15 nm from its nominal operating wavelength. Maintaining a low temperature necessitates careful thermal management of the system. We present measurements and calculations of thermal load for this design, and characterize the performance of 633 nm and 638 nm laser diodes in the low temperature regime. The laser system could be used for broad tuning of different laser diode models at a variety of wavelengths with minor modifications.

\begin{figure*}[th]
\includegraphics[resolution=600]{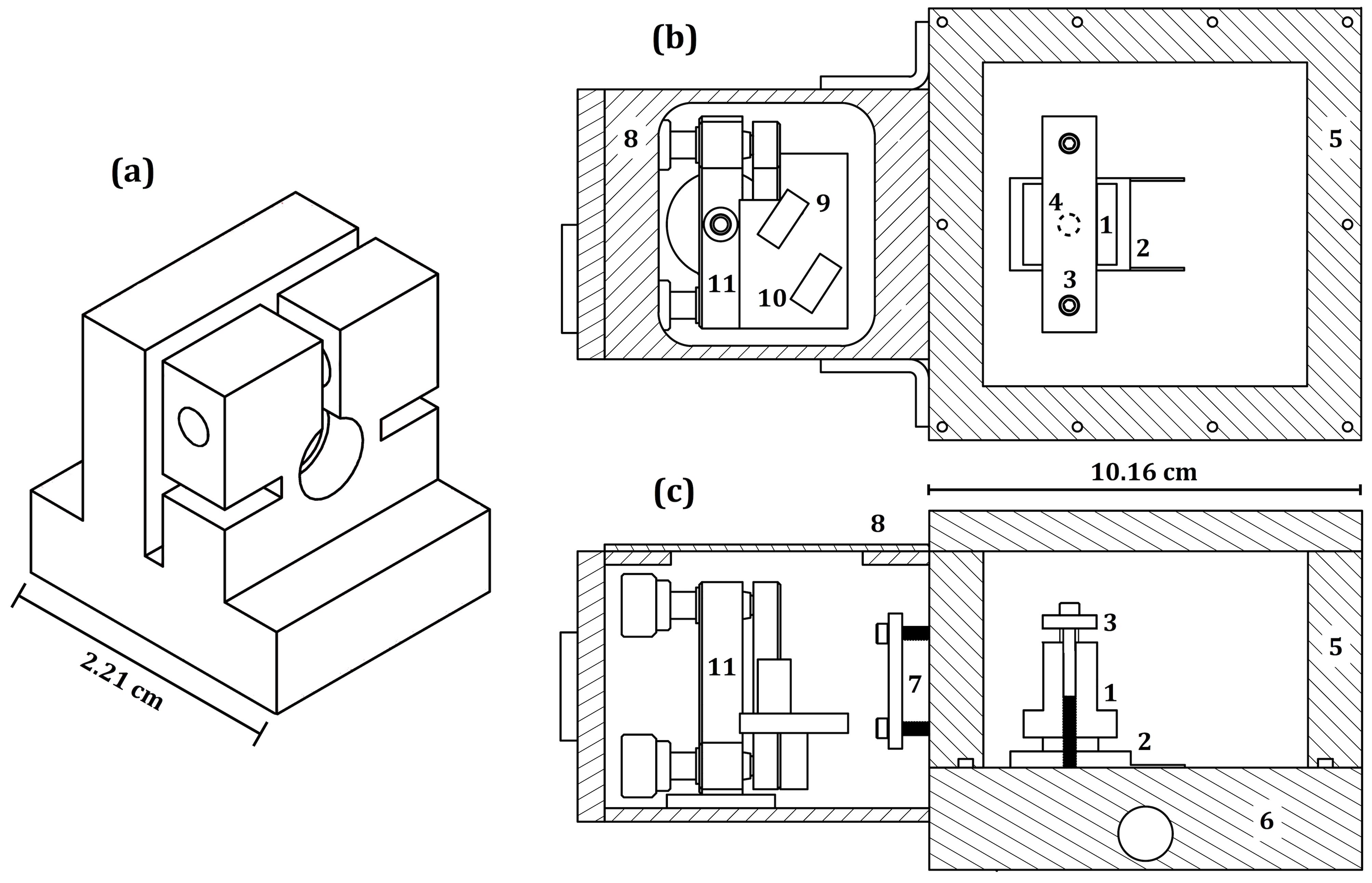}
\caption{ECDL mechanical design.
(a) Diode mount. The close face holds a TO-56 diode package and the far face holds a collimation lens. (b) Top and (c) side schematic cutout views of the low-temperature ECDL design. Important components include: 1. diode mount 2. TEC 3. clamp plate for diode mount 4. nylon spacer 5. exterior enclosure 6. water cooling block 7. front window 8. secondary enclosure 9. diffraction grating 10. mirror 11. mirror and grating mount. See text for details.
}
\end{figure*}

\section{Low-Temperature ECDL construction}

\subsection{Mechanical Design}
To achieve stable single-wavelength operation of laser diodes at the lowest possible temperature, we constructed an ECDL where the diode and optical feedback elements are housed in two separate, adjacent enclosures. Figure 1 shows a schematic of the laser system and the diode mount.

The diode is clamped inside a copper mount [Fig. 1(a), also labeled 1 on Fig. 1(b,c)] which additionally holds a collimation lens. An Omega 44008 thermistor is embedded in the diode mount for primary temperature monitoring and a Thorlabs TH10K thermistor was used to check for low-temperature measurement consistency. The diode mount sits on the cold side of a Marlow SP2402 three-stage TEC\footnote{We chose this TEC for the specified cooling performance of $82^\circ$C with a heat load of 2 W. Another TEC, the Laird Technologies MS3,119,20,15, has similar specifications.} (2) with indium foil in between. A steel clamp plate (3) presses the diode mount against the TEC from above using two threaded rods to ensure good mechanical and thermal contact. The plate itself is isolated by a 1/4'' thick nylon spacer (4, underneath clamp plate) to avoid direct thermal contact with the mount. The exterior enclosure (5) is machined from square aluminum tubing. The enclosure is vacuum-sealed by pumping air through an evacuation port on the top and can be held at low pressures to reduce convective and conductive thermal loads and to eliminate condensation. The TEC sits directly on a copper water block (6), which is cooled to 15$^\circ$C (above the dew point of $11^\circ$C) by an external water chiller (not shown) and which efficiently removes heat from the hot side of the TEC. Collimated light from the diode leaves the vacuum enclosure through a wedged window (7). All of the electrical wiring exits the box through an electrical feedthrough on the back of the enclosure (not shown). The copper block, electrical feedthrough, window, and removable enclosure top are sealed against the enclosure with Viton O-rings.

A secondary enclosure (8), at atmospheric pressure and sealed against air currents, contains a diffraction grating (9) in the Littrow configuration, with the first order diffracted beam returning to the laser diode and the zeroth order beam reflecting off a mirror (10) and exiting the enclosure through another window. The mirror and grating are positioned on a two-axis adjustable mount (11), which allows independent tuning of the grating angle in the horizontal direction for wavelength selection and vertical direction for optical feedback adjustment. Small holes on the front of the enclosure provide access to the adjustment knobs.

\subsection{Thermal Loads}

The temperature equilibrium of the cold side of the TEC is determined by the TEC operating current and the thermal load. Efficient removal of heat generated on the hot side is critical for maintaining the desired temperature difference between the hot and the cold sides of the TEC, so the design incorporates a water-cooling plate with a large heat removal capacity. The diode mount is the only object in physical contact with the cold side of the TEC. The primary thermal loads on the TEC in this arrangement are resistive heating produced by the diode as well as conduction, convection, and radiation in the system that transfer heat to the diode mount. Though precise measurement or calculation of the thermal load can be impractical, rough estimates may be obtained from basic parameters of the system\cite{Fletcher2004}.

We calculated and measured the thermal loads present in the system. For a given input electrical power ($P_I$), the laser diode produces an optical output power ($P_O$) and converts the rest into thermal load ($Q_D$) given by
\begin{align}
Q_{D}=P_I-P_O.
\end{align}

The maximum optical output power of the primary 633 nm laser diode tested was measured to be 150 mW with an input electrical power of 0.50 W, giving a maximum thermal load from the diode current of 0.35 W.

Black-body radiative thermal load ($Q_R$) from the laser enclosure onto the diode mount is given by
\begin{align}
Q_R=\epsilon\sigma A(T_E^4-T_C^4),
\end{align}
where $\epsilon$ is the emissivity of the diode mount, $\sigma$ is the Stefan-Boltzmann constant, $A$ is the surface area of the diode mount, and $T_E$ and $T_C$ are the temperatures of the enclosure and the diode mount, respectively. We measured the exposed area of the diode mount and used $\epsilon=0.1$ as an estimate for the emissivity of partially oxidized copper\cite{Roos1983}. The temperature of the enclosure (the primary source of the black-body radiation) was estimated to be $15^\circ$ C, the same as the copper water block, and the lowest diode mount temperature measured with the diode at full power was $-61.2^\circ$C. Using these parameters, the radiative thermal load was calculated to be 0.06 W, which is small compared to other sources. Placing aluminized Mylar shielding\cite{Carter1991} around the mount as a radiation shield did not have a significant effect on cooling, likely because the total radiative thermal load was already small.

Conductive thermal load ($Q_C$) is caused by heat transfer across an object connected to the diode mount and is given by
\begin{align}
Q_C=\frac{k A_C (T_H-T_C)}{L},
\end{align}
where $k$ is the conductivity of the object creating the conductive path, $A_C$ is its cross-sectional area, $L$ is its length, and $T_H$ and $T_C$ are the temperatures of the hot and cold sides of the object, respectively. The major pathway for conductive heat transfer in the system is conduction through electrical wires. The laser diode and embedded thermistor are connected to room temperature air ($22^\circ$C) through two 22 AWG and two 30 AWG wires, respectively, all with length 10 cm. The total conductive thermal load through the wires with the diode at $-61.2^\circ$C was calculated to be 0.24 W, and could be lessened by reducing the diameter of the diode wires to the minimum that could safely carry the diode current. A secondary conductive pathway is formed by the threaded rods and steel clamp plate connecting the water block ($15^\circ$C) to the diode mount ($-61.2^\circ$C). To minimize conduction, we separated the diode mount and the clamp plate with a nylon spacer. Using a direct measurement of the clamp plate temperature ($-2.9^\circ$ C) and the spacer dimensions, the conductive thermal load from this source was calculated to be 0.04 W. Conductive thermal load from heat transfer through the air was calculated to be smaller than 0.01 W.

Convective thermal load ($Q_V$) is caused by temperature gradients within the system that create air currents which transfer heat between objects with different temperatures. The convective thermal load can be written as
\begin{align}
Q_V=h A(T_A-T_C),
\end{align}
where $h$ is the convective heat transfer coefficient, $A$ is the total surface area of the diode mount exposed to convective heating, $T_A$ is the temperature of the ambient air inside the enclosure, and $T_C$ is the temperature of the diode mount. In general, $h$ depends on many variables, including the temperature difference $T_A-T_C$, the specific geometry of the diode and the enclosure, and temperature- and pressure-dependent properties of the air\cite{Incropera2007}.

To study the thermal loads from the air in the system, the diode temperature was measured as a function of pressure (Fig. 2). 
\begin{figure}
\includegraphics[resolution=600]{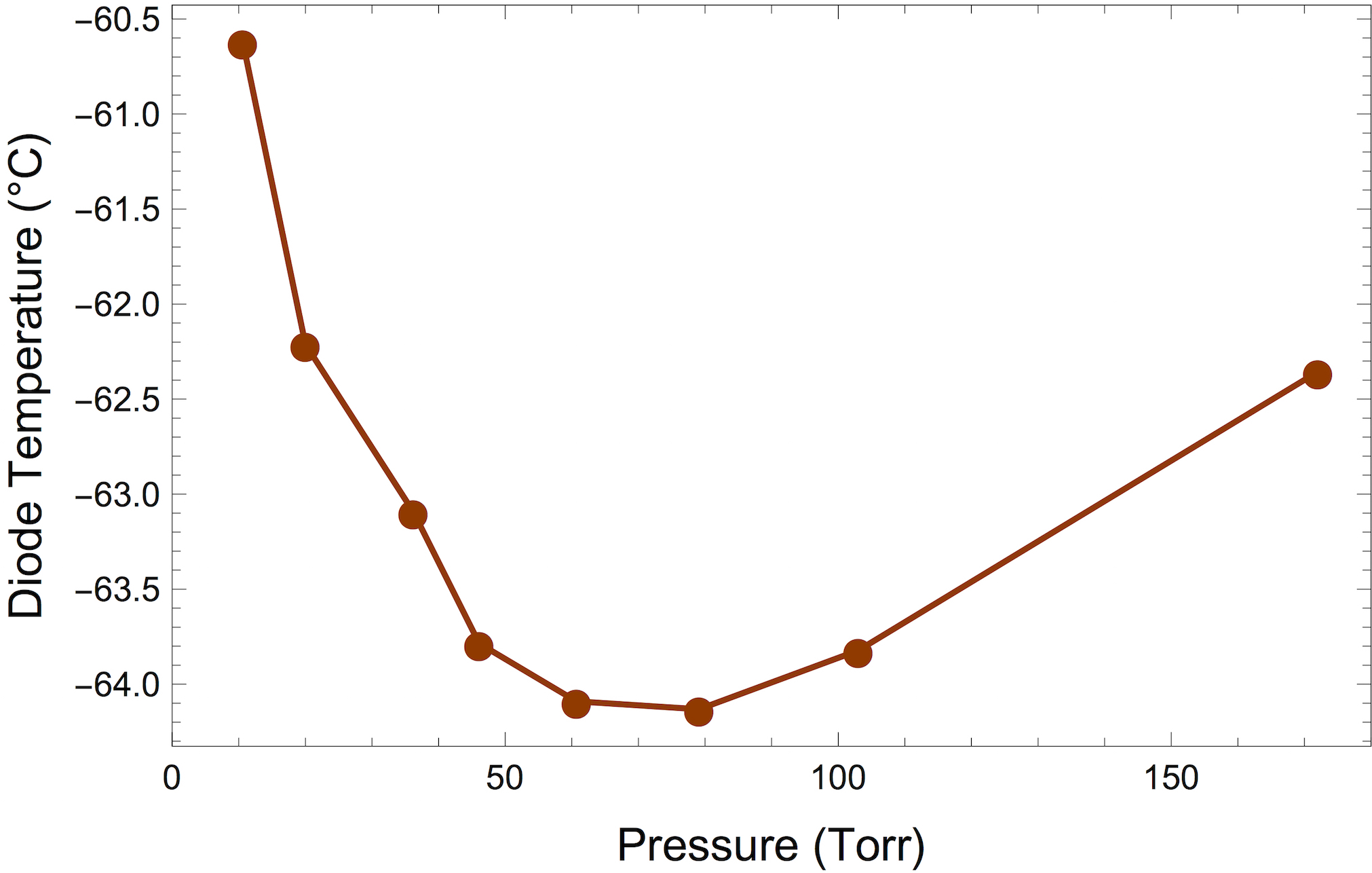}
\caption{
Lowest diode mount temperature reached as a function of air pressure in the enclosure, measured with the laser diode off. The minimum was reached between 60-80 Torr.}
\end{figure}
A lower diode temperature indicates a smaller thermal load. We found the diode temperature reached a minimum between 60-80 Torr, so all further measurements were conducted near 70 Torr. We note that convective and conductive thermal loads in general strictly decrease with decreased pressure\cite{Lemmon2004}. However, we theorize that the specific geometry of the different elements in the system tested may lead to competition between detrimental convection, which increases the thermal load on the TEC, and beneficial convection, which aids in heat dissipation from the hot side of the TEC. Although the mechanism is not understood, our finding conveniently favored operation at ``low vacuum.''

As a simplified model for calculating the convective thermal load present in the laser system, the diode mount was treated as a set of four vertical plates [Fig. 1(a)], each with a length and width of 0.02 m. Convective thermal load on the horizontal faces of the diode mount was neglected. Detailed expressions for convective thermal load on vertical plates and in systems with many other geometries can be found in Ref. 18. The temperature of the air involved in convection inside the enclosure was assumed to be between $15^\circ$C (the temperature of the water cooling block) and $22^\circ$C (air temperature outside the enclosure). To experimentally verify the model for convection, we compensated the reduction in thermal load between 760 Torr and 70 Torr using power dissipated in an adjustable thermal load\textemdash a resistor in a copper mount in place of the diode mount\textemdash which could be directly measured. Figure 3 shows the measured thermal load difference in comparison with the calculated difference and the inset shows the calculated values of $h$ at $T_A=22^\circ$C for 760 Torr and 70 Torr.
\begin{figure}
\includegraphics[resolution=600]{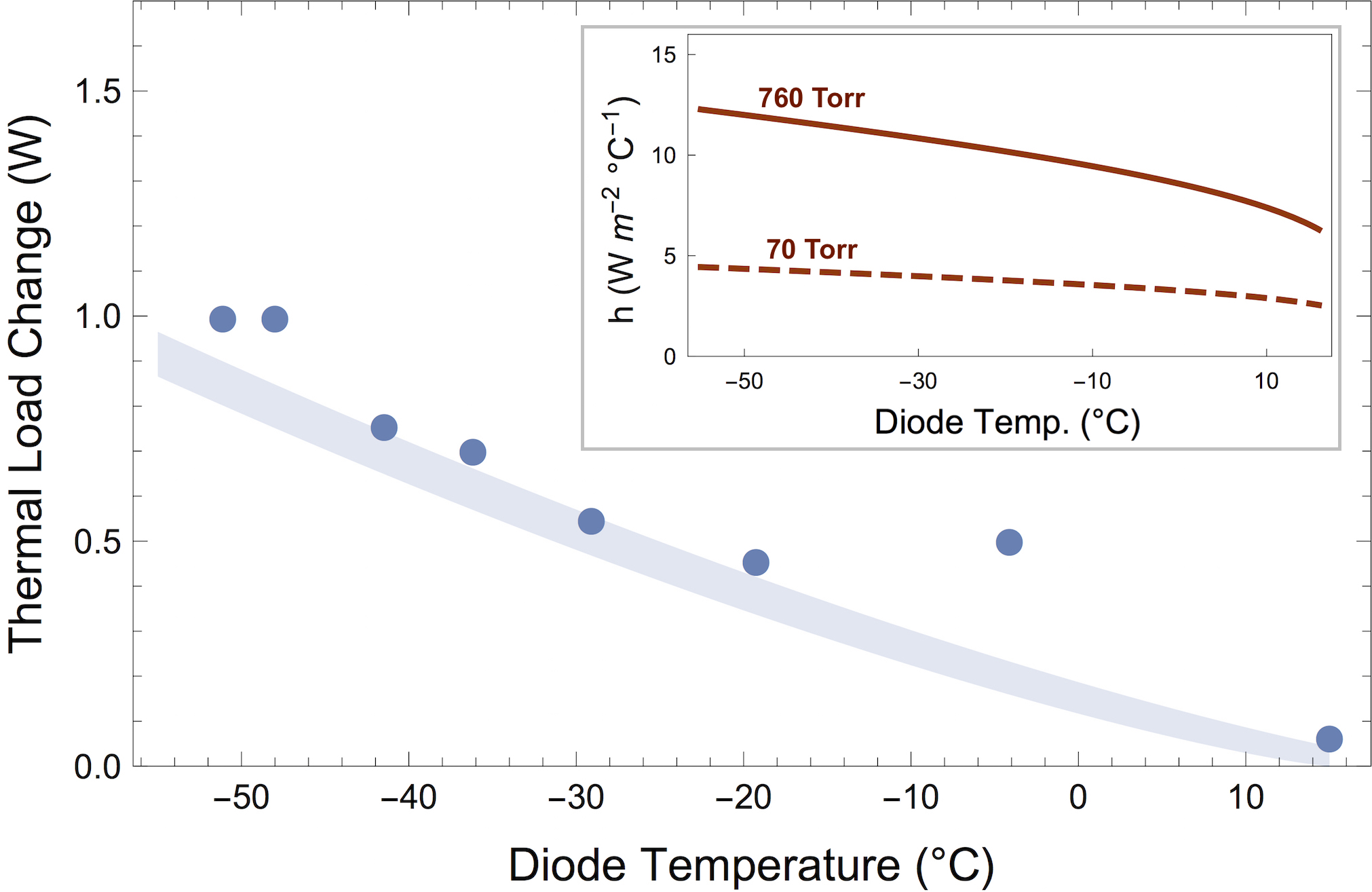}
\caption{ Convective thermal load. Main plot: thermal load change between 760 Torr and 70 Torr as a function of diode temperature. The difference (circles) was measured by introducing thermal load through resistors to compensate the change in thermal load between the two pressures tested. The measured values were compared to calculated values (shaded area) assuming air temperatures inside the enclosure between $15^\circ$C and $22^\circ$C, using the method described in the text and further in Ref. 18. Inset: the convective heat transfer coefficient $h$ calculated at 760 Torr (solid line) and 70 Torr (dashed line) as a function of diode temperature.}
\end{figure}
With no free parameters, the model and experiment produced similar results for the reduction in thermal load over the entire temperature range. The convective thermal load was calculated to be 1.68 W at 760 Torr and 0.60 W at 70 Torr, the largest thermal load in the system.

The contributions from all thermal load sources, with the diode temperature at $-61.2^\circ$C and 150 mW of optical output power, are summarized in Table I.
\begin{table}
\caption{Thermal loads in the system. The thermal load from the diode is directly measured and the other thermal loads are calculated from the parameters of the system. The measurements and calculations of thermal loads are for a 633 nm diode at $-61.2^\circ$C outputting 150 mW of optical power.}
\begin{tabular}{| l | c |}
\hline
\textbf{Source} & \textbf{Thermal Load (W)} \\
\hline
Convection (760 Torr)&1.68\\
Convection reduction (at 70 Torr) & $-1.08$\\
Diode& 0.35\\
Conduction (wires) & 0.24\\
Radiation& 0.06\\
Conduction (other) & 0.04\\
\hline
Total & 1.29\\
\hline
\end{tabular}
\end{table}
The performance of the TEC with a given thermal load can be quantified by the maximum temperature difference achieved between the cold and hot sides. As more current is supplied to the TEC, the section of the cooling block directly underneath the hot side of the TEC increases in temperature from $15^\circ$C, changing the overall temperature difference. We measured the temperature of the cooling block at distances from the center of the TEC of 18.9 mm, 23.9 mm, and 28.4 mm at a variety of electrical currents supplied to the TEC. The temperature gradient measured along the cooling block at maximum TEC power was $0.028 ^\circ$C/mm, from which we extrapolated the temperature of the water block directly under the TEC to be $17.7^\circ$C. Accounting for the heating of the water block, the total temperature difference between the diode mount and the hot side of the TEC on the water block was $78.9^\circ$C.

\section{Low-Temperature Laser Diode Performance}
Two models of laser diodes were cooled and characterized: the Oclaro HL63163DG, with a nominal center wavelength of 633 nm and a maximum output power of 100 mW, and the Oclaro HL63263DG, with a nominal center wavelength of 638 nm and a maximum output power of 200 mW (all specifications given for $25^\circ$C). Two laser diodes of each model were tested, herein termed 633-1, 633-2 and 638-1, 638-2.

Figure 4 summarizes the wavelength dependence on temperature for all four laser diodes tested.
\begin{figure}
\includegraphics[resolution=600]{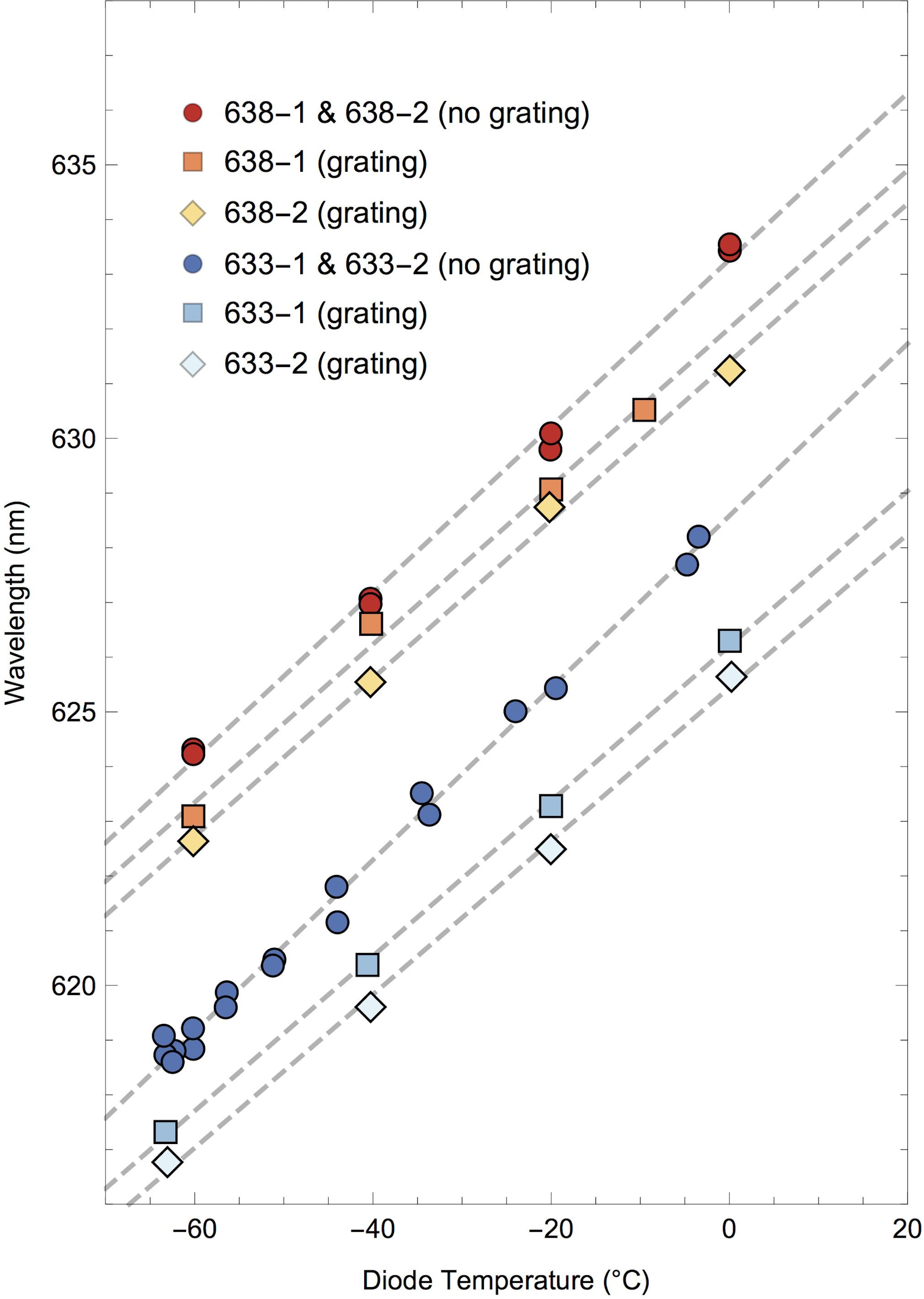}
\caption{
Diode wavelength  tuning performance at low temperature. Two diode models are tested. The circular data points represent free-running (no grating) wavelength measurements, and the square and diamond data points represent single-mode wavelengths recorded with the grating feedback. The dashed lines are linear fits, indicating tuning coefficients between 0.14-0.16 nm/$^\circ$C.
}
\end{figure}
For each diode and for the temperature range between $0^\circ$C and $-64^\circ$C, spectra were taken both free-running (without grating feedback) and with the grating pulling the wavelength down. All wavelength measurements were recorded at the maximum laser diode current where single-mode operation was achievable. Single-mode operation was achieved for each diode, with a linewidth within the minimum wavemeter resolution (Bristol 721A, resolution 2 GHz).

With the grating, all four diodes tested had a linear relationship between temperature and wavelength with a tuning coefficient between 0.14-0.15 nm/$^\circ$C. Without the grating, the tuning coefficient was between 0.15-0.16 nm/$^{\circ}$C. The temperature tunability of laser diode wavelengths is highly dependent on the type of laser diode being used. For example, a prior study of ECDL temperature tuning of a 785 nm diode observed a decrease in wavelength of over 0.2 nm/$^{\circ}$C\cite{Fletcher2004}.

The minimum wavelengths achieved with diodes 633-1 and 633-2 were 617.3 nm and 616.8 nm, respectively, a wavelength difference of approximately 16 nm from the design wavelength. The output power at the shortest wavelength achieved with the grating was 38 mW. The output power with the grating was about 16\% lower than the free-running power. With the diffraction gratings, the minimum wavelengths of the 633-1 and 633-2 diodes at $0^\circ$C were 2.5 nm and 3.2 nm below the free-running wavelengths, respectively. The diodes could not be pulled as far at the lowest temperature.

Similar wavelength changes were measured with the 638 nm laser diodes, which intrinsically produce higher optical power. The minimum wavelengths achieved with diodes 638-1 and 638-2 were 623.1 nm and 622.6 nm, respectively, approximately 15 nm below the design wavelength. The output power at the shortest wavelength was 69 mW, significantly higher than the optical power of the 633 nm diodes at the expense of longer wavelength. At 0$^{\circ}$C, diodes 638-1 and 638-2 could be pulled 1.2 nm and 1.8 nm, respectively, from their free-running center wavelengths.

The output power of a laser diode increases with decreasing temperature\cite{Wieman1991,Bohdan2008}. Figure 5 shows the relationship between output power, current, and temperature for diode 633-1 without the grating. 
\begin{figure}
\includegraphics[resolution=600]{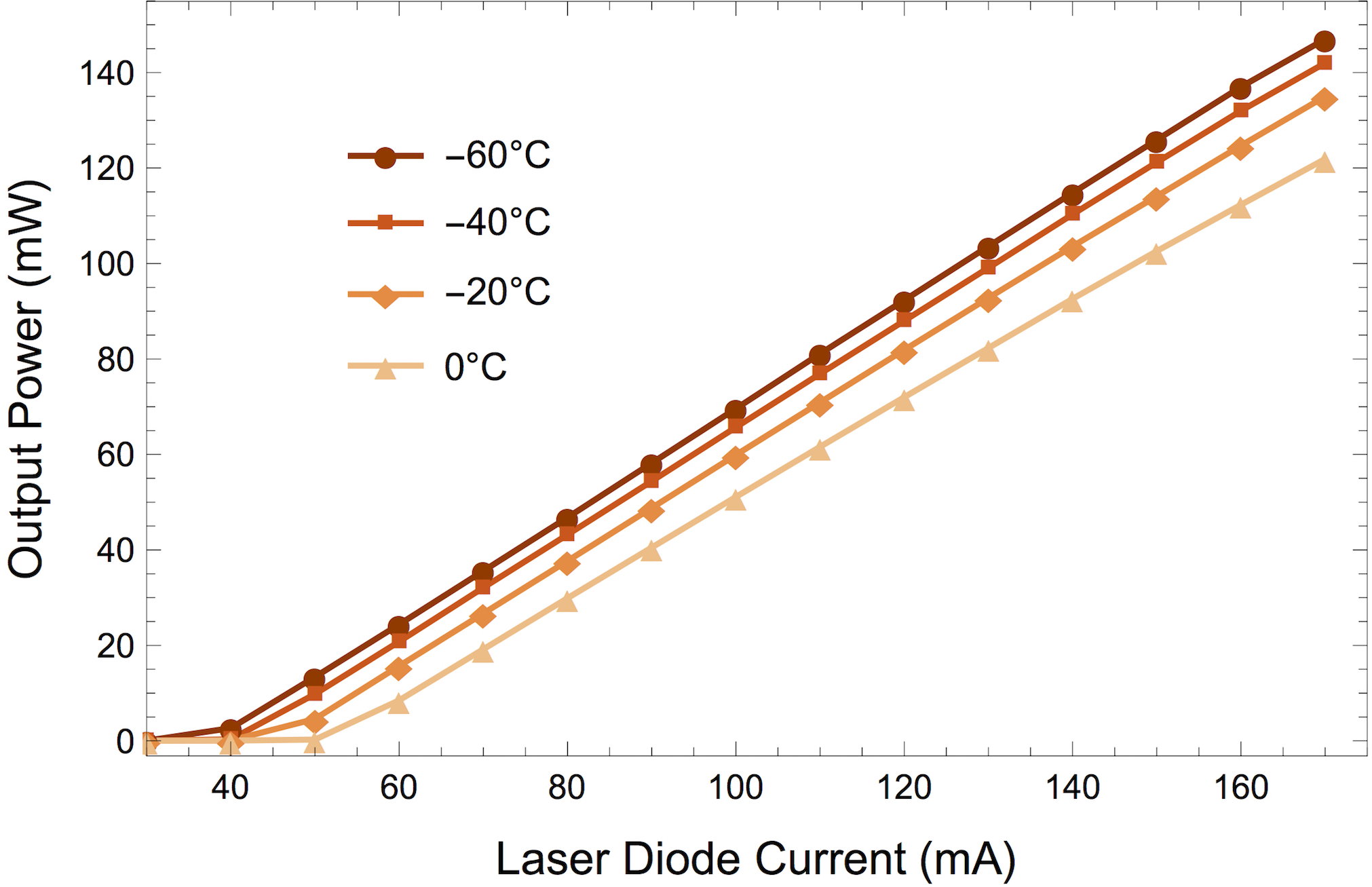}
\caption{
Laser free-running output power vs. current for different temperatures, for diode 633-1. The output power increased with decreasing temperature.
}
\end{figure}
\begin{figure}
\includegraphics[resolution=600]{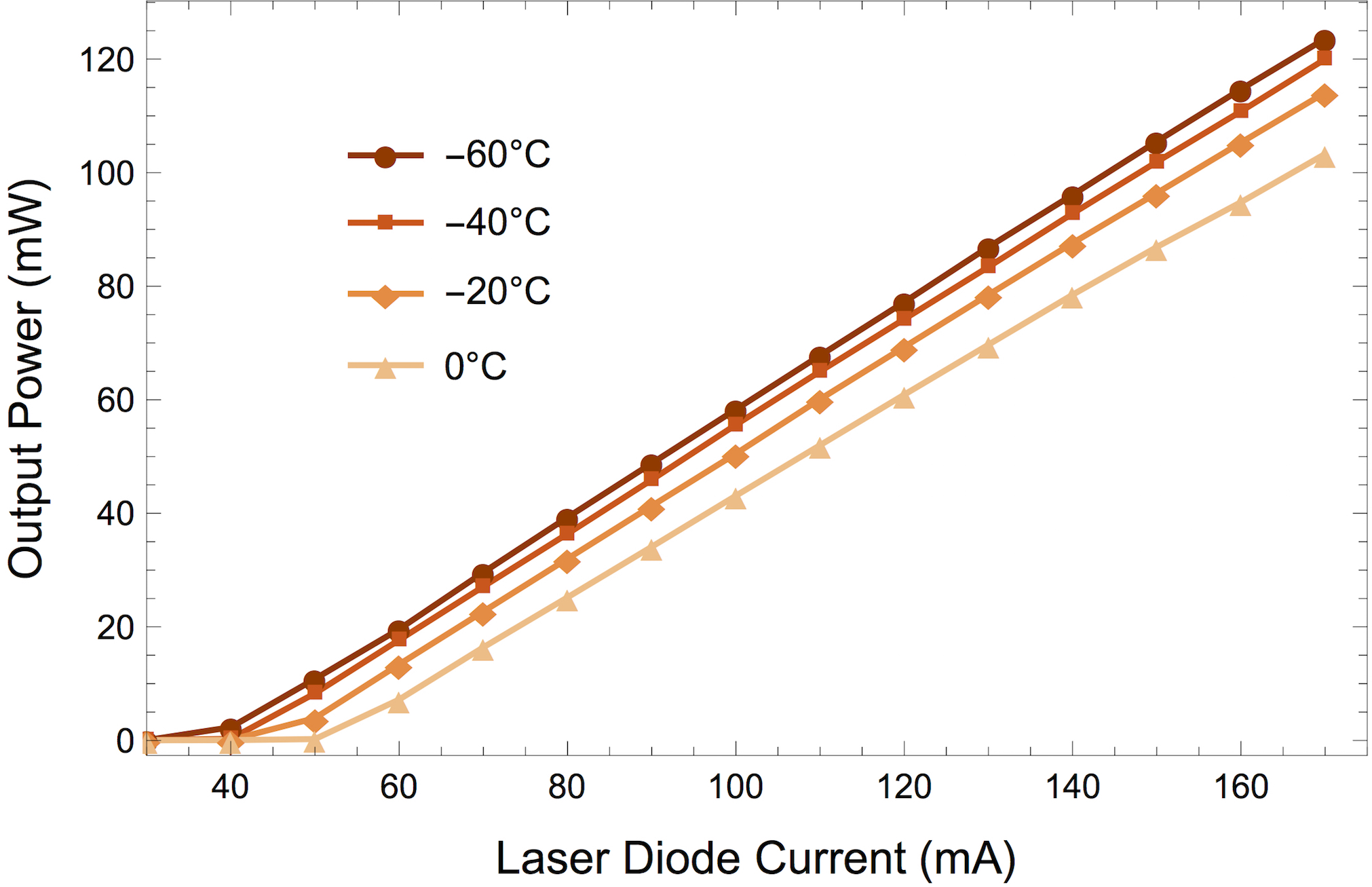}
\caption{
Laser output power vs. current for different temperatures, with secondary enclosure attached, for diode 633-1. The output power increased with decreased temperature. Output powers with the secondary enclosure were approximately 16\% lower than output powers without the enclosure, due to power lost to feedback and reflection.
}
\end{figure}
Figure 6 shows similar data for the same diode with the diffraction grating attached.
With and without the grating, the maximum power achieved was 123.6 mW and 147 mW, respectively, much higher than the output power of 100 mW at $25^\circ$C specified by the manufacturer. Furthermore, the lasing threshold current decreased with decreasing temperature, going from the specified value of 70 mA input current at $25^\circ$C to 51 mA at 0$^{\circ}$C to 38 mA at $-60^{\circ}$C.

\section{Conclusion}

We have demonstrated a design for a temperature-tunable external cavity diode laser capable of operation $85^\circ$C below ambient temperature. At the lowest temperature, the ECDL system achieved single-mode operation with two different diode models at more than 15 nm below their design wavelengths with moderate power. Both diode models output significantly higher power at low temperatures than specified at room temperature. The laser system is compact, stable, and can be primarily constructed with commercial parts along with custom parts requiring straightforward machining.

The wavelengths and laser output powers achieved in this experiment enable a range of experiments involving molecular transitions with wavelengths between 617 nm and 633 nm. The diode cooling system could be reconfigured for most other diode wavelengths and packages. 

In future work, the cooling system could be further improved while maintaining the overall simplicity of the design. A different TEC model with higher cooling capacity \footnote{For example, the Marlow SP2724 four-stage TEC, which could cool the TEC by $110^\circ$C at a heat load of 2 W but at a significantly increased price} would reduce the diode temperature for a given thermal load. The laser diode mount could be soldered directly to the TEC instead of clamped on top, removing the conductive pathway through the steel clamp plate and facilitating heat transfer between the cold face of the TEC and the mount. The wires carrying current to the laser diode could be reduced in diameter, lowering the conductive thermal load through this pathway. Finally, depending on the geometry of the interior of the enclosure, convective thermal load could be eliminated by lowering the pressure of the system by several orders of magnitude. Improvements in the minimum diode temperature would allow operation at shorter wavelengths and would further broaden the applications of the ECDL system presented here.

\section{Acknowledgements}

We thank Lee R. Liu, Yichao Yu, and John M. Doyle for helpful discussions. W. G. T. acknowledges support from the Harvard Quantum Optics Center. This work is supported by the NSF through the Harvard-MIT Center for Ultracold Atoms and the Arnold and Mabel Beckman Foundation.\\

\bibliography{Bibliography}
\end{document}